\documentclass[doublecol]{epl2} 

\title{Structural Properties of Stiff Elastic Networks}
\shorttitle{Title} 

\author{G. Gurtner \and M. Durand}

\institute{Laboratoire Mati\`{e}re et Syst\`{e}mes Complexes (MSC) \\UMR 7057 CNRS \& Universit\'{e}
Paris Diderot \\10 rue Alice Domon et L\'{e}onie Duquet, 75205 Paris Cedex 13, France, EU}

\pacs{6.70.-p}{Application of continuum mechanics to structures}

\usepackage{amsfonts}
\usepackage{amsmath}
\usepackage{amssymb}
\usepackage{graphicx}
\usepackage{tabularx}%

\abstract{
Networks of elastic beams can deform either by stretching or bending of their
members. The primary mode of deformation (bending or stretching) crucially depends on the specific details of the network architecture. In order to shed light on the relationship between microscopic geometry and macroscopic mechanics, we characterize the structural features of networks which deform uniformly, through the stretching of the beams only. We provide a convenient set of geometrical criteria to identify such networks, and derive the values of their effective elastic moduli. The analysis of these criteria elucidates the variability of mechanical response of elastic networks. In particular, our study rationalizes the difference in mechanical behavior of cellular and fiber networks.}

\begin{document}

\maketitle

\section{Introduction}
Various elastic systems can be understood as networks of interconnected rods
which deform by a combination of bending, stretching, twisting and shearing
mechanisms. Examples include polymer gels, protein networks and cytoskeletal
structures \cite{Thorpe,Head,Head2,Wilhelm,Heussinger,Heussinger3,Buxton}, crystal atomic
lattices and granular materials \cite{Limat,Ostojic}, paper \cite{Cox,Ostoja}, wood, foams, and
bones \cite{Gibson,Ashby,Deshpande,Roberts,Christensen}, and even continuous elastic bodies under certain circumstances \cite{Hrennikoff}.
Moreover, the pairwise interaction potentials used in standard elastic percolation models can also be identified with the strain energy of elastic beams \cite{Thorpe,Limat,Sahimi}.  
Despite extensive research
\cite{Head,Head2,Wilhelm,Heussinger,Heussinger3,Buxton,Dunlop,Kabla,Onck,Onck2}, the connection between the mechanical
properties of such networks on a macroscopic level and the description of
their structures on a microscopic level has not been completely elucidated
yet.
Interestingly, under identical loading conditions, some structures appear
to deform primarily through the local stretching of the beams, while in other
structures the elastic energy is stored via local bending \cite{Gibson,Ashby}
(twisting and shearing contributions are usually neglected). For instance, ``foam-like" cellular architectures tend to be bending-dominated \cite{Heussinger,Heussinger3}, while fibrous architectures exhibit a rich mechanical behavior: 
Head \textit{et al.} \cite{Head,Head2} and Wilhelm and Frey \cite{Wilhelm} simulated the two-dimensional elastic deformation of a network of cross-linked fibers and observed a transition from a nonaffine, bending-dominated regime to an affine, stretch-dominated regime with increasing density of fibers. Recent experimental studies \cite{Gardel,Liu} and mean field theories \cite{Das,Heussinger2} have confirmed this transition. Buxton and Clarke \cite{Buxton} also characterized a similar bending-to-stretching transition in three-dimensional networks, in terms of the connectivity of nodes.
Elucidating this variability of mechanical response is of interest to structural applications, as well as to our understanding of various biological systems. 

With this aim in view, we analyze in this letter the structural conditions under which a network of beams deforms uniformly (affinely) through the extension or compression of its members. 
Only some specific network geometries are compatible with such an affine, stretch-dominated, deformation.
Indeed, the network architecture must meet two requirements:
the possible symmetries of the structure, and the mechanical equilibrium at every point of the network, respectively. The first requirement results in restrictions on the beam angular distribution. We will limit our analysis to isotropic structures, though the reasoning can be transposed without difficulty to materials with lower symmetries. The second requirement results in restrictions on the possible configurations of the junctions.
The inspection of these requirements provides a convenient set of geometrical criteria to identify the structures that deform affinely.
Moreover, the analysis of these geometrical criteria rationalizes the observations reported on the mechanical behavior of cellular and fiber networks, and clarifies how the microscopic structural parameters (density of beams, density and connectivity of nodes,...) affect the macroscopic mechanical response of such networks.
We will consider essentially athermal systems, and specify how the results can be extended to the case of thermal beams.
\section{Isotropy requirement}
\begin{figure}[h]
\includegraphics[width=8cm]{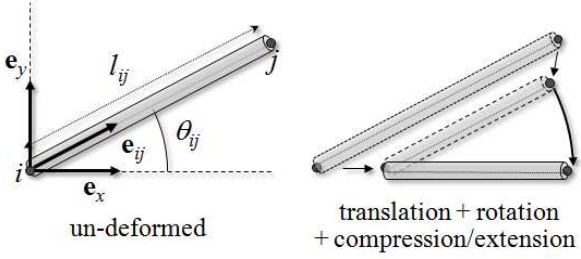}
\caption{Geometry of a typical beam
$(i,j)$. Under an affine displacement field, the beam is subjected to a combination of translation, rotation, and compression/extension.}%
\label{notations}%
\end{figure}

Affine strain, being a combination of translation, rotation and extension, induces only stretching and compression of beams (see Fig. \ref{notations}).
Therefore, an analytic expression for the elastic energy associated with such a strain can be easily derived.
Let $\epsilon_{ij}$ be the relative change in length of the beam linking nodes $i$ and $j$. Then, the elastic energy of this beam is $\kappa_{s}l_{ij}\epsilon_{ij}^{2}/2$, where $l_{ij}$ is its initial length, and $\kappa_{s}$ the one-dimensional stretch/compression modulus. $\kappa_{s}$ is determined by the Young's modulus of the beam material $E_{0}$ and the beam cross-sectional area $s$: $\kappa_{s}=E_{0}s$.  
For slender rods ($\sqrt{s} \ll l_{ij}$), the contribution of nodes to the strain energy can be neglected in comparison with the stretching energy of beams. The energy of the whole network is then simply obtained by summing over all the beams that are under extension/contraction: 
\begin{equation}
\mathcal{E}=\frac{\kappa_{s}}{2}\sum_{(i,j)}l_{ij}\epsilon_{ij}^{2}.
\label{eq:bigenergy}
\end{equation}
For an affine strain, the displacement field $\mathbf{u}$ is a linear application of the position $\mathbf{r}$: $\mathbf{u}=\mathbf{A}\cdot\mathbf{r}$, where $\mathbf{A}$ is the matrix associated with the strain.
For instance, $A_{\alpha\beta}=A_{0}\delta_{\alpha x}\delta_{\beta y}$ for a uniform shear strain $A_{0}$ in the $xy$ plane, and $A_{\alpha\beta}=A_{0}\delta_{\alpha\beta}$ for a uniform radial strain $A_{0}$. For small strains, $\epsilon_{ij}=\mathbf{e}_{ij}\cdot\left(\mathbf{u}_{j}-\mathbf{u}_{i}\right)/l_{ij}$, where $\mathbf{u}_{j}-\mathbf{u}_{i}$ is the relative displacement of nodes $i$ and $j$, and $\mathbf{e}_{ij}$ is the unit vector pointing from $i$ to $j$. Therefore, the relative change in length can be written as 
\begin{equation}
\epsilon_{ij}=\sum_{\alpha,\beta} A_{\alpha\beta} e_{ij}^{\alpha} e_{ij}^{\beta},
\label{eq:extension}
\end{equation}
where $e_{ij}^{\alpha}=$ $\mathbf{e}_{\alpha}\cdot\mathbf{e}_{ij}$ is the
cosine of the angle between the beam $(i,j)$ and the $\alpha$ axis ($\alpha\in\{x,y\}$ for two-dimensional
materials, and $\alpha\in\{x,y,z\}$ for three-dimensional materials).
Usually, isotropic networks are idealized as continuous and uniform angular distributions of identical beams \cite{Head,Head2,Christensen,Roberts}. 
However, this simplistic model does not account for the structural limitations imposed by the isotropic symmetry.
Actually, isotropic networks might contain beams with different lengths, or distributed non-uniformly.
By definition, the strain energy of an isotropic network must be invariant under rotations and reflections of the strain field. Therefore, the energy expression deduced from Eqs (\ref{eq:bigenergy}) and (\ref{eq:extension}) must be invariant under the substitution $\mathbf{A} \rightarrow \mathbf{R^{T}} \cdot \mathbf{A} \cdot \mathbf{R}$, for any orthogonal matrix $\mathbf{R}$. After a little algebra, the application of these invariance properties leads to the following set of relations:
\begin{equation}%
\begin{cases}
\left\langle {e_{ij}^{\alpha}} ^{2}\right\rangle =\frac{1}{d} \qquad
\left\langle {e_{ij}^{\alpha}} ^{4}\right\rangle =\frac{3}{d\left(d+2\right)} & \\
& \\
\left\langle {e_{ij}^{\alpha}} ^{2} e_{ij}^{\beta} e_{ij}^{\gamma} \right\rangle =0 \quad\left(  \beta\neq\gamma\right)  &
\end{cases}
\label{eq:cond_tot}%
\end{equation}
with $d=2$ for two-dimensional (2D) networks, and $d=3$ for three-dimensional (3D) networks.
The angular brackets denote an average over the network,
defined for any quantity $q_{ij}$ as:
$\left\langle q_{ij}\right\rangle =\sum_{(i,j)}l_{ij}q_{ij}/\sum_{(i,j)}l_{ij}$.
If the network contains beams with free end(s), they must be excluded from the above summations, since they do not contribute to its mechanical properties.
The ``isotropy conditions" (\ref{eq:cond_tot}) constitute a set of 4 (resp. 14) equations for 2D (resp. 3D) networks. As expected, these conditions are satisfied for a continuous and uniform angular distribution of identical beams. But they are also satisfied for networks with discrete angular distributions or heterogeneous beams, as it is the case for most cellular structures \cite{Gurtner}.

Networks that deform in an affine (stretch-dominated) way are stiffer than other networks of similar density. 
With the help of the isotropy conditions (\ref{eq:cond_tot}), it is straightforward to derive analytic expressions for the elastic moduli of such stiff networks. Indeed, these conditions imply that
$\left\langle (e_{ij}^{\beta} e_{ij}^{\gamma}) ^2 \right\rangle =1/\left(  d\left(d+2\right)  \right)$ and $\left\langle e_{ij}^{\beta} e_{ij}^{\gamma} \right\rangle =0$ (with $\beta\neq\gamma$).
Thus, the density of strain energy $\varepsilon$ simplifies to
\begin{equation}
\varepsilon=\frac{\kappa_{s}\rho_{c}}{d\left(d+2\right)}\left(\frac{1}{2}(\sum_{\alpha}a_{\alpha \alpha})^{2}+\sum_{\alpha, \beta}a_{\alpha \beta}^{2}\right),
\label{eq:affineenergy}
\end{equation}
with $a_{\alpha\beta}=\left( A_{\alpha\beta}+A_{\beta\alpha} \right)/2 $. $\rho_{c}$ is the corrected line density, defined as the total beam length per unit area (2D) or unit volume (3D), agreeing that only beams connected at both ends are taken into account (\textit{i.e.}: dangling ends are disregarded). Eq. (\ref{eq:affineenergy}) must be compared with the general expression for the strain energy density of an isotropic body in linear elasticity \cite{Landau}:
\begin{equation}
\varepsilon=\frac{\lambda}{2}(\sum_{\alpha}u_{\alpha\alpha})^{2}+\mu\sum_{\alpha,\beta}u_{\alpha\beta}^{2},
\label{eq:energy}%
\end{equation}
where $\lambda$ is the Lam\'{e}'s first parameter, $\mu$ the shear modulus (or
Lam\'{e}'s second parameter), and $u_{\alpha\beta}=\frac{1}{2}\left(  \frac{\partial
u_{\alpha}}{\partial x_{\beta}}+\frac{\partial u_{\beta}}{\partial x_{\alpha}}\right)$ are the
components of the strain tensor ($u_{\alpha}$ are the components of the
displacement field).
One obtains for the effective Lam\'{e}'s parameters of an isotropic stiff network:%
\begin{equation}
\lambda=\mu=\frac{\kappa_{s}}{d\left(d+2\right)}\rho_{c}.
\end{equation}
Any other elastic modulus of an isotropic body is related to $\lambda$ and $\mu$ \cite{Landau},
and hence can be easily evaluated. Values of most common moduli are reported
on table \ref{tableau}.
\begin{table}[ptb]
\centering
\newcolumntype{Y}{>{\centering\arraybackslash}X}
\begin{tabularx}{\linewidth}[c]{|Y|Y|Y|Y|Y|Y|}\hline
& $\lambda/\kappa_{s}$ & $\mu/\kappa_{s}$ & $E/\kappa_{s}$  &  $K/\kappa_{s}$ & $\nu$\\\hline
\textbf{2D} & $\rho_c /8$ & $\rho_c /8$ & $\rho_c /3$  &  $\rho_c /4$ & $1/3$\\\hline
\textbf{3D} & $\rho_c /15$ & $\rho_c /15$ & $\rho_c /6$  &  $\rho_c /9$ & $1/4$\\\hline
\end{tabularx}
\caption{Elastic moduli of stiff isotropic networks normalized by the
one-dimensional stretching modulus $\kappa_{s}$: Lam\'{e}'s first parameter ($\lambda$), shear modulus ($\mu$), Young's modulus ($E$), bulk modulus ($K$), and Poisson's ratio ($\nu$).}%
\label{tableau}%
\end{table}
These values coincide with those reported in the literature \cite{Head,Head2,Torquato,Roberts,Gibson,Christensen,Durand3} (note, however, that our values of $E$ and $\nu$ correct those given by Head \textit{et al.} \cite{Head,Head2}). 
They constitute upper-bounds for the macroscopic moduli of networks with similar density $\rho_c$: any deviations from affine deformation can only lower the stiffness of the material \cite{Gurtner}.
It can also be noticed that these elastic moduli vary linearly with $\rho_{c}$, in agreement with scaling arguments \cite{Buxton,Ashby}.
\section{Mechanical equilibrium requirement}
Obviously, the isotropy conditions (\ref{eq:cond_tot}) alone are not sufficient to identify the structures which deform in an affine manner: the compatibility of an affine deformation with the equations of mechanical equilibrium must also be inspected.
The forces and torques acting on any junction of the network must balance at equilibrium. The (tensile) force exerted
by the beam $\left(i,j\right)$ on the junction $i$ is
$\kappa_{s}\epsilon_{ij}\mathbf{e}_{ij}$, where $\epsilon_{ij}$ is related to the affine strain field by Eq. (\ref{eq:extension}). Thus, the moment of this axial force is zero, and the mechanical equilibrium conditions reduce to the force balance: $\sum_{j}\epsilon_{ij}\mathbf{e}_{ij}=\mathbf{0}$, where the summation is over all the nodes that are connected to the node $i$ (agreeing that dangling ends are disregarded). This equality must hold for any orientation of the strain field. Using the same rotational invariance argument as for the energy expression, we obtain a set of structural conditions at every junction $i$ of the 2D (resp. 3D) network, which can be summarized as:
\begin{equation}
\sum_{j} e_{ij}^{\alpha} e_{ij}^{\beta} e_{ij}^{\gamma} =0,
\label{eq:cond_forces}%
\end{equation}
for all $\alpha$, $\beta$, $\gamma\in\left\{x,y\right\}$ (resp. $\left\{x,y,z\right\}$). Therefore, the mechanical equilibrium requirements lead to a set of 4 (resp. 10) equations per node for 2D (resp. 3D) networks. 
These conditions, along with the isotropy conditions (\ref{eq:cond_tot}), constitute a set of necessary and sufficient conditions for an affine, stretch-dominated, deformation.

The mechanical conditions (\ref{eq:cond_forces}) impose severe restrictions on the geometry and valency of a junction.
Some of the implications of these conditions are analyzed below for 2D networks. Let us note $z_{i}$ the connectivity of node $i$ (number of beams connected to it, with the exception of dangling ends), and $\theta_{ij}$ the angle between the beam $(i,j)$ and the $x$ axis. Introducing the complex variables $y_{ij}=\exp \left(  \imath \theta_{ij} \right) $,
the mechanical equilibrium conditions (\ref{eq:cond_forces}) simplify to 
\begin{align}
\sum_{j=1}^{z_{i}} y_{ij}=0, & & \sum_{j=1}^{z_{i}} y_{ij}^3=0.
\label{eq:cos_et_sin}
\end{align}
Clearly, there is no solution to this set of equations for two- ($z_{i}$=2) and three-fold ($z_{i}$=3) junctions
(configurations with beams all collinear are left out).
In agreement with Maxwell's criterion \cite{Dunlop, Deshpande, Thorpe3, Wyart, Heussinger2, Kellomaki}, we conclude that a stiff network must have a node connectivity equal to or greater than 4 (for large structures). Furthermore, it can be shown \cite{Gurtner} that the only possible configurations for a four-fold junction ($z_{i}$=4) are those with beams parallel in pairs: $\theta_{i3}=\theta_{i1}+\pi$, $\theta_{i4}=\theta_{i2}+\pi$\footnote{The analysis of the solutions for nodes with valency $z_{i}\geq 5$ is more delicate. Nodes with an even number of adjoining beams have trivial solutions: beams parallel in pairs. But it is not clear whether these are the only solutions.}. 
These results shed light on the transition observed in computational and experimental studies of cross-linked fibers \cite{Head,Head2,Wilhelm,Liu}: because of the finite fiber length, there are two-, three-, and four-fold coordinated cross-links in such networks. By construction, the mechanical conditions (\ref{eq:cos_et_sin}) are fulfilled at the four-fold junctions, but not at the two- and three-fold junctions. At low density of fibers ($\rho_{c}L \sim 1$, where $L$ is the length of a fiber), there is a significant number of two- and three-fold junctions, so the network deforms in a non-affine way\footnote{\label{note3}Such a network will deform through floppy modes if the typical elastic energy of a node is much smaller than the typical bending energy of a beam (free hinges)\cite{Kellomaki}, and primarily through bending modes in the opposite limiting case (fixed angles).}. When the density of fibers increases, the proportion of two- and three-fold nodes decreases and the proportion of four-fold nodes increases, leading asymptotically ($\rho_{c}L \rightarrow \infty$) to an affine strain regime\footnote{The random orientation of fibers ensures that isotropy conditions (\ref{eq:cond_tot}) are satisfied.}.
This is consistent with earlier observations \cite{Head,Head2,Wilhelm,Liu} reporting that the deformation of a fiber network becomes asymptotically affine as the number of cross-links per fiber ($\sim \rho_{c}L$) increases. 
It must be noted that this analysis is valid for small strains only. At larger strains, even low-density fiber networks will eventually become stretch-dominated \cite{Kabla,Onck,Onck2}. 

Similarly, our results reveal why cellular networks are almost always
bending-dominated \cite{Heussinger,Heussinger3,Deshpande}: such structures usually do not meet the mechanical conditions (\ref{eq:cond_forces}). This is specifically the case of open-cell foams.
Foams are particular cellular materials: their structures
result from a surface minimization process, leading to geometrical and
topological rules, known as Plateau's laws.  Unlike a closed-cell foam, the cell walls of an open-cell foam disappear during the drying process, leaving only a network of interconnected edges. Plateau's laws state that these edges meet in threefold
(resp.\ fourfold) junctions with equal angles of  $120^\circ$
(resp. $109.5^\circ$) in 2D (resp. 3D) foams. Such node configurations
cannot satisfy the equilibrium conditions (\ref{eq:cond_forces}). Therefore, open-cell foams will
rather deform by bending of their edges.

\begin{figure}[h]
\includegraphics[width=8cm]{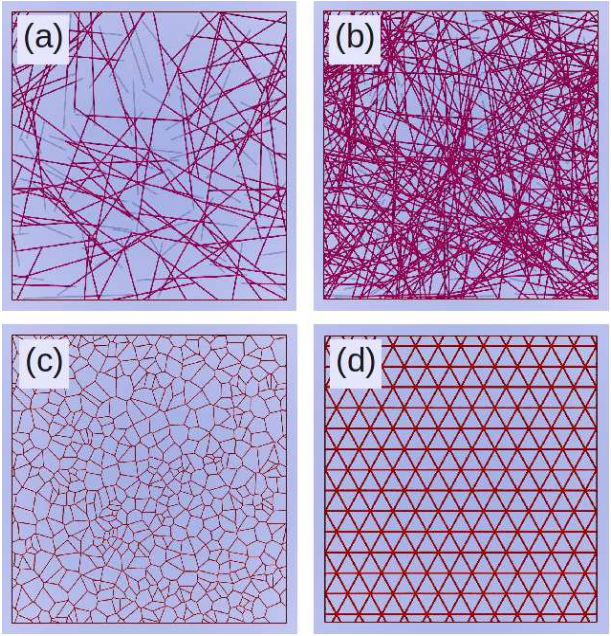}
\caption{Examples of two-dimensional networks: low-density fiber network (a), high-density fiber network (b), Voronoi diagram of a random set of points in the plane (c), and triangular lattice (d). For convenience, dangling ends in the fiber networks are shaded off. By construction, the low-density fiber network (a) and Voronoi diagram (c) contain nodes with valency $< 4$, and so deform in a non-affine way (see footnote \ref{note3}). On the other hand, the high-density fiber network (b) and the triangular lattice (d) satisfy the mechanical conditions (\ref{eq:cond_forces}) at every node. Consistent with our analysis, these two networks deform uniformly.}%
\label{simulations}%
\end{figure}

Figure \ref{simulations} illustrates this discussion. Four different structures are depicted: two fiber networks at low and high density, respectively, and two cellular networks: the Voronoi diagram of a random set of points in the plane, and a triangular lattice. For each of these structures, the isotropy conditions (\ref{eq:cond_tot}) are met. But only the fiber network in the limit of high density (rigorously, when $\rho_{c}L \rightarrow \infty$) and the triangular lattice satisfy the mechanical conditions (\ref{eq:cond_forces}) at every node, and thus deform uniformly.
Incidentally, it is worth noting that our results apply to both random and regular structures. The triangular lattice is one example of regular structure deforming in an affine way. Other examples of two- and three-dimensional regular structures are given in \cite{Gurtner}. 

In the above, we have considered a purely athermal model of networks. However, many biological systems are comprised of thermally fluctuating polymers. In these systems, the effects of temperature on the elastic properties of the polymer are quantified by the persistence length $l_p$, defined as the ratio of bending stiffness to thermal energy $l_p = \kappa_{b}/(k_B T)$ ($\kappa_{b}=E_{0}s^{2}/(4 \pi)$ for a cylindrical beam). Such thermal fluctuations result in an effective, length-dependent, stretch modulus \cite{Head,Head2,Heussinger3}: $\kappa_{s}^{th} \sim \kappa_{b} l_p / l_{ij}^{3}$ for a thermal beam of length $l_{ij}$. This entropic compliance dominates for long enough beams, giving rise to a distinct affine regime. Accordingly, the mechanical condition (\ref{eq:cond_forces}) becomes for this affine entropic regime: $\sum_{j} e_{ij}^{\alpha} e_{ij}^{\beta} e_{ij}^{\gamma}/l_{ij}^{3} =0$. Unlike in the affine mechanical regime, the mechanical conditions in the affine entropic regime depend on the beam length distribution.
It must be mentioned that an alternative model has been recently proposed to explain the mechanical response of biological networks, in which the elastic network is composed of rigid rods connected by flexible cross-linkers \cite{Broedersz,Kasza}.
\section{Conclusion}
In summary, we analyzed the structural features of isotropic networks which deform uniformly through the stretching/compression of their beams.
The study of these structural features sheds light on the relationship between the structural details of a network and its macroscopic mechanical behavior, and rationalizes the variability of mechanical response of diverse elastic networks (fibrous and cellular, regular and disordered). In particular, our analysis confirms the previously observed bending-to-stretching transition in fiber networks. The elastic moduli of networks that deform affinely are also derived, and can be simply expressed in terms of the beam stretch modulus and the line density. We hope these
results are of interest to structural applications, as well as to our understanding of biological systems. 

\acknowledgments
We thank A. J. M. Cornelissen, A.\ Emeriau and C. Gay for critical reading of the manuscript.


\begin{thebibliography}{99}                                                                                               %

\bibitem {Thorpe}M.\ F.\ Thorpe, \textit{Phys. Biol.} \textbf{4} 60--63 (2007).

\bibitem {Heussinger}C.\ Heussinger and E.\ Frey, \textit{Phys.\ Rev.\ Lett.}
\textbf{96}, 017802 (2006).

\bibitem {Heussinger3}C.\ Heussinger and E.\ Frey, \textit{Phys.\ Rev.\ E}
\textbf{75}, 011917 (2007).

\bibitem {Head}D.\ A.\ Head, A.\ J.\ Levine, and F.\ C.\ MacKintosh,
\textit{Phys.\ Rev.\ Lett.} \textbf{91}, 108102 (2003).

\bibitem {Head2}D.\ A.\ Head, A.\ J.\ Levine, and F.\ C.\ MacKintosh,
\textit{Phys.\ Rev.\ E.} \textbf{68}, 061907 (2003).

\bibitem {Wilhelm}J. Wilhelm and E.\ Frey, \textit{Phys.\ Rev.\ Lett.}
\textbf{91}, 108103 (2003).

\bibitem {Buxton}G.\ A.\ Buxton and N.\ Clarke, \textit{Phys.\ Rev.\ Lett.}
\textbf{98}, 238103 (2007).

\bibitem {Limat}L.\ Limat, \textit{Phys. Rev. B} \textbf{40}, 9253-9268 (1989).

\bibitem {Ostojic}S. Ostojic and D. Panja, \textit{Phys. Rev. Lett.}
\textbf{97}, 208001 (2006).

\bibitem {Cox}H. L. Cox, \textit{Br. J. Appl. Phys.}
\textbf{3}, 72-79 (1952).

\bibitem {Ostoja}M. Ostoja-Starzewski and D. C. Stahl, \textit{J. Elasticity} \textbf{60}, 131-149 (2000).

\bibitem {Gibson}L. J. Gibson and M. F. Ashby, Cellular Solids - Structure and
properties, Cambridge Univ. Press (1997, 2$^{nd}$ edition).

\bibitem {Ashby}M.F.\ Ashby, \textit{Philosophical transactions of the Royal
Society A} \textbf{364}, 15-30 (2006).

\bibitem {Deshpande}V.\ S. Deshpande, M. F. Ashby and N.\ A.\ Fleck,
\textit{Acta mater.} \textbf{49}, 1035-1040 (2001).

\bibitem {Roberts}A. P. Roberts and E. J. Garboczi, \textit{J. Mech. Phys.
Solids} \textbf{50}, 33-55 (2002).

\bibitem {Christensen}R. M. Christensen, \textit{J. Mech. Phys. Sol.}
\textbf{34}, 563-578 (1986). R. M. Christensen, \textit{Int. J. Solids Structs}
\textbf{37}, 93-104 (2000).

\bibitem{Hrennikoff} A. Hrennikov, \textit{J. Appl. Mech.} \textbf{8}, 169-175 (1941).

\bibitem {Sahimi}M. Sahimi, \textit{Phys. Rep.} \textbf{306}, 213-395 (1998).

\bibitem {Dunlop}J. Dunlop, W. Richard, P. Fratzl, and Y. Br\'{e}chet, http://hdl.handle.net/2042/15749

\bibitem {Kabla}A. Kabla and L. Mahadevan, \textit{J. R. Soc. Interface}
\textbf{4}, 99-106 (2007).

\bibitem {Onck}P. R. Onck, T. Koeman, T. van Dillen, and E. van der Giessen, \textit{Phys. Rev. Lett.}
\textbf{95}, 178102 (2005).

\bibitem {Onck2}E. M. Huisman, T. van Dillen, P. R. Onck, and E. Van der Giessen, \textit{Phys. Rev. Lett.}
\textbf{99}, 208103 (2007).

\bibitem {Gardel}M. L. Gardel, J. H. Shin, F. C. MacKintosh, L. Mahadevan, P. Matsudaira, and D. A. Weitz, \textit{Science} \textbf{304}, 1301-1305 (2004).


\bibitem {Liu}J. Liu, G. H. Koenderink, K. E. Kasza, F. C. MacKintosh, and D. A. Weitz, \textit{Phys. Rev. Lett.} \textbf{98}, 198304 (2007).

\bibitem {Das}M. Das, F. C. MacKintosh, and A. J. Levine \textit{Phys.\ Rev.\ Lett.}
\textbf{99}, 038101 (2007).

\bibitem {Heussinger2}C. Heussinger, B. Schaefer, and E.\ Frey, \textit{Phys. Rev. E}
\textbf{76}, 031906 (2007).


\bibitem {Landau}L.\ Landau and E.\ Lifchitz, Theory of elasticity, Pergamon
Press, New York (1986).



\bibitem {Torquato}S.\ Torquato, L.\ V. Gibiansky, M.\ J. Silva and L.\ J.
Gibson, \textit{Int. J. Mech. Sci.} \textbf{40}, 71-82 (1998).


\bibitem {Durand3}M.\ Durand, \textit{Phys. Rev. E.} \textbf{72}, 011114 (2005).


\bibitem {Gurtner}G.\ Gurtner and M.\ Durand, \textit{to be published} (2009).


\bibitem {Thorpe3}D. J. Jacobs and M. F. Thorpe, \textit{Phys. Rev. E}
\textbf{53}, 3682-3693 (1996).

\bibitem {Wyart}M. Wyart, H. Liang, A. Kabla, and L. Mahadevan, \textit{Phys. Rev. Lett.}
\textbf{101}, 215501 (2008).

\bibitem{Kellomaki} M. Kellom\"aki, J. \AA str\"om, and J. Timonen \textit{Phys. Rev. Lett.}
\textbf{77}, 2730-2733 (1996).

\bibitem{Broedersz} C. P. Broedersz, C. Storm, and F. C. MacKintosh, \textit{Phys. Rev. Lett.}
\textbf{101}, 118103 (2008).

\bibitem{Kasza} K.E. Kasza, G. H. Koenderink, Y. C. Lin, C. P. Broedersz, W. Messner, F. Nakamura, T. P. Stossel, F. C. MacKintosh, and D. A. Weitz, \textit{Phys. Rev. E} \textbf{79}, 041928 (2009).







\end{thebibliography}
\end{document}